\documentclass{article}
\usepackage[preprint]{spconf}
\usepackage{amsmath,graphicx, multirow, cite}
\usepackage[pdftex]{}

\usepackage[hyphens]{url}
\usepackage{hyperref}
\usepackage[hyphenbreaks]{breakurl}

\copyrightnotice{\copyright\ IEEE 2021}
\toappear{Accepted as a conference paper at ICASSP 2021, Toronto, Canada
}

\title{Fast DCTTS: Efficient Deep Convolutional Text-to-Speech}

\name{ Minsu Kang
            \textsuperscript{1}, 
        Jihyun Lee
            \sthanks{currently, in KAIST graduate school.}\textsuperscript{2}, 
        Simin Kim
            \textsuperscript{3} and
        Injung Kim
            \textsuperscript{4}
}

\address{
    Handong Global University, Pohang, Republic of Korea\\ \\
    \{\textsuperscript{1}mskang, \textsuperscript{3}21400126, \textsuperscript{4}ijkim\}@handong.edu, \textsuperscript{2}jyun.lee@kaist.ac.kr
}

\begin{document}
\ninept
\maketitle
  
\begin{abstract}
    We propose an end-to-end speech synthesizer, Fast DCTTS, that synthesizes speech in real time on a single CPU thread. The proposed model is composed of a carefully-tuned lightweight network designed by applying multiple network reduction and fidelity improvement techniques. In addition, we propose a novel \textit{group highway activation} that can compromise between computational efficiency and the regularization effect of the gating mechanism. As well, we introduce a new metric called \textit{elastic mel-cepstral distortion (EMCD)} to measure the fidelity of the output mel-spectrogram. In experiments, we analyze the effect of the acceleration techniques on speed and speech quality. Compared with the baseline model, the proposed model exhibits improved MOS from 2.62 to 2.74 with only 1.76\% computation and 2.75\% parameters. The speed on a single CPU thread was improved by 7.45 times, which is fast enough to produce mel-spectrogram in real time without GPU.  
\end{abstract}


\section{Introduction}
    \label{sec:intro}
    In recent few years, deep-learning based speech synthesis technology has been rapidly improved. Tacotron\cite{Tacotron} is the end-to-end speech synthesis model that has demonstrated the possibility of the neural speech synthesizer. The following work, Tacotron2\cite{Tacotron2} exhibits a fidelity comparable to that of human speech. Tacotron model consists of the encoder-attention-decoder structure based on recurrent neural networks (RNN), which requires a long time to learn. To accelerate training speed, Tachibana et al. propose a TTS model, called DCTTS\cite{DCTTS}, that is based on convolutions instead of recurrences. Leveraging parallel computing, the training speed of DCTTS is faster than that of Tacotron, while the quality of the output speech is comparable. Transformer-TTS\cite{Transformer-TTS} is based on Transformer\cite{Transformer} network which has shown successful results in many Natural Language Processing (NLP) tasks. Transformer-TTS\cite{Transformer-TTS} exhibits slightly improved speech quality compared with Tacotron2\cite{Tacotron2}.

    In regard to synthesis speed on a high-performance GPU, the most important bottleneck of the aforementioned models is the autoregressive decoding process that requires sequential computations. Recent works including FastSpeech\cite{FastSpeech}, AlignTTS\cite{AlignTTS}, and FastSpeech2\cite{FastSpeech2} apply non-autoregressive architecture and significantly improve synthesis speed. However, their improvements are achieved mainly by parallelization rather than reduction in computations.
    
    In this paper, we propose a neural TTS model that synthesizes high-fidelity mel-spectrogram in real time without a GPU. We designed a lightweight network by applying multiple techniques to reduce computation and improve fidelity. In addition, we propose a novel \textit{group highway activation} that can compromise between computational efficiency and the regularization effect of the gating mechanism. For the evaluation of speech quality, we propose a new metric called \textit{elastic mel-cepstral distortion (EMCD)} that measures the distance between the synthesized mel-spectrogram and the ground truth considering skipping and repeating errors.
            
    In experiments, we evaluate the effect of the acceleration techniques measuring their influence on fidelity using EMCD and mean opinion score (MOS). The experimental results demonstrate that the network reduction techniques are effective in accelerating a neural TTS model. Compared with the baseline model, the proposed model exhibits improved MOS from 2.62 to 2.74 with only 1.76\% computation and 2.75\% parameters\footnote{Sound demos are available at: \url{https://jackson-kang.github.io/paper_works/FastDCTTS/}}. The synthesis speed on a single CPU thread was improved by 7.45 times, which is fast enough to produce mel-spectrogram in real time without a GPU.
    
    On the other hand, our experimental results suggest that the depthwise separable convolution, which is known effective in the image recognition field, does not improve the speed of neural TTS in practice, although it significantly reduces the computation of 1D convolution in theory. We have also observed that replacing the highway activation of DCTTS with residual connection degrades the output quality significantly.
    
    The contributions of our work are summarized as follows$\colon$
    \begin{itemize}
        \item We present a neural TTS model that generates a mel-spectrogram in real time on a single CPU thread.
        \item We qualitatively evaluate the effect of multiple techniques to reduce computation and improve fidelity on a neural TTS model using EMCD.
        \item We propose a novel group highway activation that can compromise between computational efficiency and the regularization effect of the gating mechanism.
        \item We propose a new evaluation metric, EMCD, to measure the quality of a mel-spectrogram.
    \end{itemize}
   
    \let\thefootnote\relax\footnote{\\© 2021 IEEE. Personal use of this material is permitted.  Permission from IEEE must be obtained for all other uses, in any current or future media, including reprinting/republishing this material for advertising or promotional purposes, creating new collective works, for resale or redistribution to servers or lists, or reuse of any copyrighted component of this work in other works.}
    
\section{Related Works}
    \subsection{Baseline model}
    \label{sec:related_works}
    We chose DCTTS \cite{DCTTS} as the baseline model because of two primary reasons: first, there are many acceleration techniques applicable for CNNs\cite{Mobilenet, Conv-pruning, Deepvoice3, Shufflenet} most of which have been developed for image recognition. Second, DCTTS \cite{DCTTS} can be trained and evaluated in a short time. In practice, this is an important advantage when searching for the best model through a large number of experiments.
    
    The Text2Mel network of DCTTS \cite{DCTTS} is composed of a text encoder, an attention module, an audio encoder, and an audio decoder. The text encoder converts the input text into two sequences of character embeddings, $K$ and $V$. The audio encoder converts the previously generated mel-spectrogram into a sequence of audio embeddings $Q$. The attention module computes an attention matrix $A$ from $K$ and $Q$, which represents the alignment between the parts of the input text and the output mel-spectrogram. The audio decoder generates the mel-spectrogram of the next time step from $A$, $V$ and $Q$ in an auto-regressive way. 
    
    The number of layers in the text encoder, audio encoder and audio decoder are 14, 13, and 11, respectively. All networks apply the highway activation\cite{Highwaynetwork}. Additionally, we applied weight normalization\cite{Weightnorm} to improve training efficiency.

    \subsection {Building lightweight networks}
    Neural TTS models require a huge amount of computation. Most of the successful results in TTS acceleration have been achieved through parallelization. However, to improve synthesis speed on a CPU, we need to reduce the actual amount of computation.
    
    In the image processing field, various techniques have been developed to reduce the computational requirement of CNN. Iandola et al. present SqueezeNet\cite{Squeezenet}, a lightweight network composed of Fire modules. MobileNet\cite{Mobilenet} reduces the amount of computation by applying depthwise separable convolution, width multiplier, and resolution multiplier. ShuffleNet\cite{Shufflenet} improves computational efficiency by applying group convolution combined with channel shuffling. MobileNet2\cite{Mobilenetv2} further improves MobileNet\cite{Mobilenet} by applying additional techniques such as linear bottlenecks and inverted residuals. Another popular technique to build a reduced network is network pruning\cite{Pruning} that removes less important parts of a neural network to save computation and parameters.

 \section{Optimization Techniques for Neural TTS}
    \subsection{Computational optimization techniques}
        \label{sec:optimization_neural_tts}
        \subsubsection{Depthwise separable convolution}
            Depthwise separable convolution\cite{Mobilenet} in image processing fields approximates a 3D convolution ($channel \times height \times width$) with a 2D depthwise convolution ($height \times width$) and a 1D pointwise convolution ($channel$). In speech synthesis, depthwise separable convolution \cite{Mobilenet} decomposes 2D convolution ($channel \times time$) into a 1D depthwise convolution and a 1D pointwise convolution. The computational complexity of 2D convolution is $O(D_K M N D_F)$, where $D_K$ and $D_F$ are the sizes of the convolution filter and the feature map. $M$ and $N$ denote and the number of the input and output channels, respectively. On the other hand, the complexity of depthwise and pointwise convolutions are $O(D_K M D_F)$ and $O(M D_F D_F)$, respectively. Consequently, in theory, the reduction in computation is $\frac{1}{N} + \frac{1}{D_K}$.
            
        \subsubsection{Group highway activation}
            \begin{equation}
                \centering
                \begin{aligned}
                    y=T(x, W_{T})H(x, W_{H}) + C(x, W_{C})x
                \end{aligned}
                \label{equation:highwayactivation}
            \end{equation}
            DCTTS \cite{DCTTS} applies highway activation layers\cite{Highwaynetwork} as equation (\ref{equation:highwayactivation}), where $x$ and $y$ denote the input and output feature maps, respectively, and $H(x, W_{H})$ is an operator such as convolution. $T(x, W_{T})$ and $C(x, W_{C})$ are the transformation and carry gates.  $W_{T}$, $W_{C}$, and $W_{H}$ are the parameters of the gates and operation, respectively. $C(x,W_C)$ can be replaced by $1-T(x,W_T)$. The two gates help the learning of deep neural networks. However, they increase the  computational burden by 2x or 3x, as the computational cost of each gate is the same as that of $H(x,W_H)$.
            
            To reduce computation, we applied two alternatives of highway activation \cite{Highwaynetwork}. One is residual connection\cite{Resnet}, which is a special case of highway activation \cite{Highwaynetwork} with $T(x,W_T)=C(x,W_C)=1$. Residual connection is also known effective in training deep neural networks, while requiring only minimal additional overhead. First, we replaced highway activation\cite{Highwaynetwork} with residual connection\cite{Resnet} on all layers. The resulting model is called Residual DCTTS. However, as displayed in Table \ref{table:group_highway_activation}, Residual DCTTS degrades speech quality significantly. Particularly, it substantially increases skipping and repeating errors when applied to audio encoder or audio decoder. However, it does not degrade speech quality seriously when applied only to text encoder. We presume that removing the learned gating mechanism increases noise in the intermediate acoustic features, and consequently, destabilizes attention. 
            
            After observing that the learned gating mechanism is important for attention stability, we developed the other alternative called group highway activation. The new activation combines feature elements into groups, and the elements within the group share gate values. The size of the gate vector is reduced by $\frac{1}{g}$, where $g$ is group size. Replacing highway activation by group highway activation, the computation of a convolution layer is reduced by $(1+\frac{1}{g}) / 2$. By adjusting $g$, we can search for a trade-off between computational efficiency and the regularizing effect of the gating mechanism.
            
        \subsubsection{Network size reduction}
            Reducing network size can improve synthesis speed, as well. The text encoder, audio encoder and audio decoder networks of the baseline model consist of 14, 13, and 11 convolution layers, respectively. Each layer of the audio encoder and audio decoder networks contain 256 channels. The layers of the text encoder contain 512 channels. We searched for the smallest network that maintains acceptable fidelity by reducing the number of layers and channels monitoring the quality of the output speech.
            
        \subsubsection{Network pruning}
            Network pruning\cite{Pruning} is a technique to reduce the size and computation of the neural network. After training, the pruning algorithm estimates the importance value of each unit, and then, removes the units with low importance values. The pruning of a CNN model\cite{Conv-pruning} is performed for the convolution filters or feature maps. In this study, we applied the pruning method of \cite{Conv-pruning} with a few modification to fit group highway activation. First, we remove less important feature maps. Then, we remove the gate elements whose features have been removed.

    \subsection{Fidelity improvement techniques}
        \label{sec:fidelity_improvement_techniques}
            Positional encoding is a way to add a piece of information that reflects absolute or relative position to the feature vector. Particularly, it improves attention stability by helping to learn the temporal relation between feature vectors\cite{Deepvoice3}. In this research, we applied the scaled positional encoding of Transformer-TTS\cite{Transformer-TTS}. Positional information is added to the feature vector as \({x'}_{i} = {x}_{i} + \alpha PE (pos, i)\), where $PE(pos, i)$ is defined as \(PE (pos, 2k) = sin ({pos \over base^{2k / dim}})\) for $i=2k$, and \(PE(pos, 2k + 1) = cos ({pos \over base^{2k / dim}})\) for $i=2k+1$. $\alpha$ is a trainable weight.
            
            In addition to positional encoding, we applied scheduled sampling\cite{Scheduled-sampling}. In the training of an autoregressive decoder, each step requires the output of the previous step, which makes it difficult to parallelize. Applying teacher forcing, we can train the autoregressive decoder using the corresponding part of the ground-truth mel-spectrogram instead of the output of the previous step. However, teacher forcing cause a discrepancy that the model is optimized for the ground truth in training, while it generates a new output from the previous output. Scheduled sampling\cite{Scheduled-sampling} alleviates such discrepancy. Initially, the model learns mainly from the ground truth. As the learning progresses, it increases portion of the previous output.
        
\section{Elastic Mel Cepstral Distortion (EMCD)}
    \label{sec:EMCD}
    To build a high-fidelity lightweight TTS model, we need to evaluate not only speed but also speech quality. We evaluate the quality of the output mel-spectrogram by the distance from the ground truth. However, it is not simple to measure the distance between mel-spectrograms that are not aligned. For example, the Euclidean metric does not work well for the mel-spectrograms when one of them contains the skipping or repeating problem. For this reason, we have developed a new evaluation metric, \textit{elastic mel cepstral distortion (EMCD)}, that measures the difference between mel-spectrograms considering the alignment between them.
    \begin{equation}
        \begin{aligned}
            D(i, &j) = w_m \times MCD(x_i, y_j) 
            \\ & + min \big\{D(i, j-1), D(i-1, j), D(i-1, j-1)\big\}
        \end{aligned}
        \label{equation:emcd}
    \end{equation}
    Mel cepstral distortion(MCD)\cite{MCD} is known effective in measuring perceptual distance between speech signals and expressed as $MCD(i, j) = \sqrt{2 \sum_{d=1}^{D} ({x}_{d}[i] - {y}_{d}[j])^2}$, where $i=\{1,...,{T}_{syn}\}$, $j=\{1,...,{T}_{gt}\}$, $x$ and $y$ are the sequences of mel frequency cepstral coefficient (MFCC) converted from the synthesized and ground truth mel-spectrograms, and $T_{syn}$ and $T_{gt}$ denote their length, respectively. We have extended MCD by combining it with dynamic time warping (DTW)\cite{DTW}, to measure distance while finding the optimal alignment via dynamic programming. Each step of the alignment process is formulated as equation (\ref{equation:emcd}), where $w=[w_{hor}, w_{ver}, w_{diag}]$ and $m=argmin \big\{ D(i, j-1), \ D(i-1, j), \ D(i-1, j-1)\big\}$.
    The weight vector $w$ is composed of hyper-parameters to control the penalties for the horizontal, vertical, and diagonal transitions, each of which corresponds to repeating, skipping, and matching, respectively. 
    
    In this research, we set $w=(1,1,\sqrt{2})$, which leads the matching algorithm to assign a penalty weight of $w_{diag}=\sqrt{2}$ to the diagonal matching path from $(i-1,j-1)$ to $(i,j)$ and a penalty weight of $w_{hor} + w_{ver} = 2$ to the rectangular matching trajectory between the same points. After matching, $EMCD(T_{syn}, T_{gt})$ provides EMCD between $x$ and $y$. In evaluation, we normalize the EMCD value by $T_{gt}$ to eliminate the effect of speech length.

    The idea of EMCD is similar to that of MCD-DTW\cite{MCD-DTW}. However, in contrast to MCD-DTW\cite{MCD-DTW}, EMCD assigns different penalty weights to horizontal, vertical, and diagonal transitions, and therefore, is more effective in measuring the difference caused by skipping and repeating.
    \begin{table}[]
        \centering
        \begin{tabular}{|c|c|c|c|}
            \hline
            \multirow{2}{*}{\textbf{Model}}                                      & \multirow{2}{*}{\textbf{\begin{tabular}[c]{@{}c@{}}Syn. time \\ (compared to B$_{GPU}$)\end{tabular}}} & \multicolumn{2}{c|}{\textbf{EMCD}}             \\ \cline{3-4} 
                                                                                 &                                                                                                                 & \textbf{LJ}           & \textbf{KSS}           \\ \hline
            \begin{tabular}[c]{@{}c@{}}Base$_{GPU}$ (HC)\end{tabular} & \begin{tabular}[c]{@{}c@{}}1.28 sec (1.00 $\times$)\end{tabular}                                                   & \multirow{2}{*}{9.45} & \multirow{2}{*}{10.36} \\ \cline{1-2}
            \begin{tabular}[c]{@{}c@{}}Base$_{CPU}$ (HC)\end{tabular} & \begin{tabular}[c]{@{}c@{}}6.85 sec (5.35 $\times$)\end{tabular}                                                     &                       &                        \\ \hline
            \begin{tabular}[c]{@{}c@{}}ResDCTTS\end{tabular}             & \begin{tabular}[c]{@{}c@{}}3.85 sec (3.00 $\times$)\end{tabular}                                                     & 12.93                 & 13.59                  \\ \hline
            \begin{tabular}[c]{@{}c@{}}GH DCTTS\end{tabular}        & \begin{tabular}[c]{@{}c@{}}6.37 sec (4.98 $\times$)\end{tabular}                                                     & 9.10                  & 9.29                   \\ \hline
        \end{tabular}
        \caption{Comparison of the baseline model, residual DCTTS, and group highway DCTTS. HC, ResDCTTS and GH DCTTS denote highway convolution, Residual DCTTS and Group highway DCTTS, respectively. The baseline model is denoted as Base.}
        \label{table:group_highway_activation}
    \end{table}
\section{Experiments}
    \subsection{Experimental settings}

    We used two datasets for experiments: the LJSpeech\cite{LJSpeech} and the Korean Single Speaker (KSS) speech dataset\cite{KSS}. The LJSpeech dataset\cite{LJSpeech} is composed of 13,100 audio samples, and the KSS dataset\cite{KSS} consists of 12,853 audio samples. Both datasets were recorded by a single female speaker. We used 70\% of the samples for training, and reserved the remaining portions for validation (10\%) and test (20\%) for a reliable evaluation. The total length of the training samples we used are 8.4 hours (KSS) and 16.8 hours (LJSpeech).
    
    We conducted experiments on a computer that has a E3-1240 v3 CPU (3.40 GHz) and a GTX-1080 GPU. We implemented the TTS models using PyTorch v1.1 compiled from the source code. We set batch size to one, and limited the maximum length of each mel-sequence to 200. We trained on the GPU, but measured synthesis speed on a single CPU thread.
    
    \subsection{Experimental results}   

        \subsubsection{Depthwise separable convolution}
            We reduced computation by replacing convolution with depthwise separable convolution\cite{Mobilenet}. In theoretical calculation, depthwise separable convolution\cite{Mobilenet} significantly reduces the number of operations from 275B to 100B. However, replacing ordinary convolutions by depthwise separable convolutions\cite{Mobilenet} rather increased synthesis time from 6.85 seconds to 18.16 seconds.
        
        \subsubsection{Group highway activation}
            We compared the synthesis time of the three types of activation described in section \ref{sec:optimization_neural_tts}.2 as well as their output quality using EMCD. In Residual DCTTS and Group Highway DCTTS, highway activation\cite{Highwaynetwork} was replaced by residual connections\cite{Resnet} and group highway activation (g = 2), respectively. In theory, the amount of computation of a convolution layer with group highway activation is 75\% of that of a convolution layer with highway activation. The actual synthesis times are presented in Table \ref{table:group_highway_activation}. Compared with the baseline model, Residual DCTTS almost doubled the synthesis speed, but significantly increased EMCD. Particularly, Residual DCTTS often produced mel-spectrogram containing the skipping and repeating problems. On the other hand, Group Highway DCTTS exhibited synthesis time reduced by 7\%. The actual improvement in synthesis speed was not as significant as the reduction in theoretical computation. In addition, Group Highway DCTTS reduced EMCD by 3.7\% from 9.45 to 9.10 on the LJSpeech dataset\cite{LJSpeech} and by 10.3\% from 10.36 to 9.29 on the KSS dataset\cite{KSS}.
    
        \begin{figure}
            \centering
            \includegraphics[width=\linewidth]{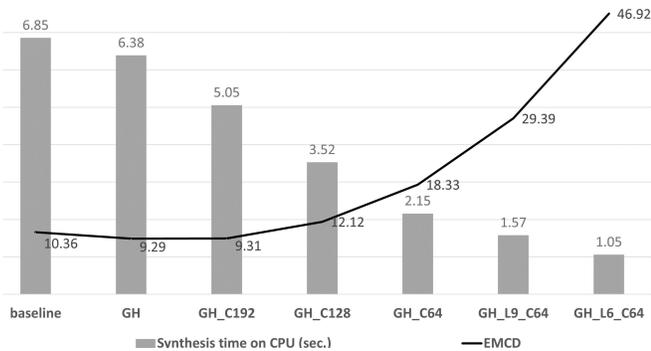}
            \caption{The effect of network size reduction on synthesis time and speech quality. The acronym GH denotes group highway activation, and the postfixes La and Cb indicate that the number of layers and channels, respectively.}
            \label{fig:network_reduction}
        \end{figure}
 
        \subsubsection{Network size reduction}
            Figure {\ref{fig:network_reduction}} displays the experimental results on KSS dataset\cite{KSS}. As expected, the smallest model, GH\_L6\_C64, was the fastest. Although it exhibited a significantly high EMCD, we chose this model for the following experiments, and applied quality improvement techniques.
            
        \begin{table}[]
            \centering
            \begin{tabular}{|c|c|c|c|}
                \hline
                \multicolumn{2}{|c|}{\multirow{2}{*}{}}                                                                                                  & \multicolumn{2}{c|}{\textbf{Weight normalization}}          \\ \cline{3-4} 
                \multicolumn{2}{|c|}{}                                                                                                                   & \textbf{on$\rightarrow{}${}on} & \textbf{on$\rightarrow{}$off} \\ \hline
                \multirow{2}{*}{\begin{tabular}[c]{@{}c@{}}LJ
                \end{tabular}} & GH\_L6\_C64                                                       & 30.59               & 15.26            \\ \cline{2-4} 
                                                                                     & \begin{tabular}[c]{@{}c@{}}GH\_L6\_C64 (10\% Pr.)\end{tabular} & Unrecognizable      & 16.24                \\ \hline
                \multirow{2}{*}{\begin{tabular}[c]{@{}c@{}}KSS
                \end{tabular}}                                                 & GH\_L6\_C64                                                       & 46.92               & 9.75                 \\ \cline{2-4} 
                                                                                     & \begin{tabular}[c]{@{}c@{}}GH\_L6\_C64 (10\% Pr.)\end{tabular} & Unrecognizable      & 10.69                \\ \hline
            \end{tabular}
            \label{effect_of_weightnorm}
            \caption{The effect of the weight normalization trick. The models marked as (10\% Pr.) are the models in which 10\% of the convolution filters were pruned.}
            \label{table:weightnorm_trick}
        \end{table}    
            
        \subsubsection{Network pruning and weight normalization trick}

            We applied network pruning\cite{Conv-pruning} to the fastest GH\_L6\_C64 model to further improve synthesis speed using KSS dataset \cite{KSS}. After pruning 10\% of the convolution filters, the synthesis time was reduced by 18.09\% from 1.05 sec. to 0.86 sec. (Note that these EMCDs and synthesis time were averaged by the number of test samples, which is 20\% of all samples.) However, the pruned models often produced unrecognizable outputs. We hypothesized that the reason was the lack of network capacity caused by the overly aggressive size reduction. To alleviate this problem, we applied a trick. To improve the speed and stability of learning, we applied weight normalization \cite{Weightnorm} at the beginning of training. However, after pruning, we deactivated weight normalization to supplement network capacity and fine-tuned the reduced network. As shown in Table \ref{table:weightnorm_trick}, the weight normalization trick remarkably decreased the EMCD on both datasets.

            Additionally, we applied the same trick to the baseline model. In this case, the trick decreased EMCD by only 1.02 and 1.44 on LJSpeech and KSS, respectively.
    
    \subsubsection{Positional encoding and scheduled sampling}
    
        We applied the two fidelity improvement techniques\cite{Transformer-TTS, Scheduled-sampling} described in Section \ref{sec:fidelity_improvement_techniques}. Positional encoding \cite{Transformer-TTS} reduced EMCD by 41.19\% from 16.24 to 9.55 on the LJSpeech dataset \cite{LJSpeech} and by 12.16\% from 10.69 to 9.39 on the KSS dataset \cite{KSS}. However, scheduled sampling\cite{Scheduled-sampling} did not lead to any improvement. 
            \begin{table}[]
                \centering 
                \begin{tabular}{|c|c|c|}
                \hline
                                                                                                           & baseline model                                                                                       & Fast DCTTS                                                                                     \\ \hline
                \begin{tabular}[c]{@{}c@{}}Text\\ Encoder\end{tabular}                                     & \begin{tabular}[c]{@{}c@{}}C-128-512\\ C-512-512\\ 12$\times$HC-512-512\end{tabular}                 & \begin{tabular}[c]{@{}c@{}}2$\times$C-128-128\\ 12$\times$RC-128-128\end{tabular}               \\ \hline
                \begin{tabular}[c]{@{}c@{}}Audio\\ Encoder\end{tabular}                                    & \begin{tabular}[c]{@{}c@{}}C-80-256\\ 2$\times$C-256-256\\ 10$\times$HC-256-256\end{tabular}              & \begin{tabular}[c]{@{}c@{}}C-80-64\\ 5$\times$GH-64-64\end{tabular}                               \\ \hline
                \begin{tabular}[c]{@{}c@{}}Audio\\ Decoder\end{tabular}                                    & \begin{tabular}[c]{@{}c@{}}C-512-256\\ 6$\times$HC-256-256\\ 3$\times$C-256-256 \\ C-256-80\end{tabular} & \begin{tabular}[c]{@{}c@{}}C-128-64\\ 4$\times$GH-64-64\\ C-64-80\end{tabular}                    \\ \hline
                \begin{tabular}[c]{@{}c@{}}Num. of\\Computations\end{tabular}                          & 275,098,419,200                                                                                 & \begin{tabular}[c]{@{}c@{}}4,835,728,000\end{tabular}                   \\ \hline
                \begin{tabular}[c]{@{}c@{}}Num. of Params\end{tabular}                            & 23,896,064                                                                                      & \begin{tabular}[c]{@{}c@{}}657,728\end{tabular}                        \\ \hline
                \begin{tabular}[c]{@{}c@{}}Syn-time$_{CPU}$\end{tabular} & \begin{tabular}[c]{@{}c@{}}6.35 sec.\end{tabular} & \begin{tabular}[c]{@{}c@{}}0.92 sec.\end{tabular} \\ \hline
                \begin{tabular}[c]{@{}c@{}}EMCD\\(LJ, KSS)\end{tabular}                            & 9.45, 10.36    & 9.55, 9.39 \\ \hline       
                \begin{tabular}[c]{@{}c@{}}MOS\\(LJ, KSS)\end{tabular}                            & 2.42, 2.62   & 2.45, 2.74   \\ \hline
            \end{tabular}
            \caption{ Comparison of DCTTS and Fast DCTTS}
            \label{table:fastDCTTS}
        \end{table}
        \subsection{Fast DCTTS}
            Table \ref{table:fastDCTTS} summarizes our best model, Fast DCTTS, compared with the baseline model, DCTTS. The proposed model requires only 1.6\% of computations and 2.75\% of parameters compared to baseline model. Its synthesis time on a single CPU thread was only only 0.92 second, which is 7.45x faster than the baseline model. It is even faster than the baseline model running on a GPU.
            
            We measured the mean opinion score (MOS) of the output speech. We used Parallel WaveGAN \cite{Parallelwavegan} for the vocoder. We randomly selected 10 samples from KSS and 10 samples from LJSpeech dataset. Then, 25 subjects rated the quality of the synthesized speeches with a score in the range of 1 (bad) to 5 (excellent). The MOS of Fast DCTTS was 2.45 on LJSpeech and 2.74 on KSS, which were slightly higher than the MOS 2.42 on LJSpeech and 2.62 on KSS of the baseline model.

    \section{Conclusion}
        \label{section:conclusion}
            We proposed a high-fidelity neural TTS, Fast DCTTS, that converts text into mel-spectrogram in real time on a single CPU thread. Its output quality is  comparable or even slightly improved compared with the baseline model DCTTS. The proposed methods reduce the actual amount of computation rather than relying on GPU-based parallel computing. Additionally, we proposed a novel group highway convolution as well as a new evaluation metric, EMCD, to measure the quality of the synthesized mel-spectrogram containing the skipping and repeating errors.
            
    \section{Acknowledgements}
        \label{section:acknowledgements}
            This work was supported by Speech AI Lab at NCSOFT Inc., and the National Program for Excellence in Software at Handong Global University (2017-0-00130) funded by the Ministry of Science and ICT.
\vfill\pagebreak

\bibliographystyle{IEEEbib}
\bibliography{strings,refs}

\end{document}